\documentclass[10pt,emptycopyrightspace]{ewsn-proc}

\usepackage{graphicx}
\usepackage{balance}
\usepackage{comment}

\numberofauthors{1}

\author{
\alignauthor Tarik Kazaz, Xianjun Jiao, Merima Kulin, Ingrid Moerman \\
    \affaddr{Department of Information Technology}\\
    \affaddr{Ghent University - imec, IDLab}\\
    \email{\{tarik.kazaz, xianjun.jiao, merima.kulin, ingrid.moerman\}@intec.ugent.be}  
}

\title{Demo: WiSCoP - Wireless Sensor Communication Prototyping Platform}

\begin{document}

\maketitle
\begin{abstract}

To enhance system performance of future heterogeneous wireless networks the co-design of PHY, MAC, and higher layer protocols is inevitable. In this work, we present WiSCoP - a novel embedded platform for experimentation, prototyping and implementation of integrated cross-layer network design approaches. WiSCoP is built on top of a Zynq hardware platform integrated with FMCOMMS1/2/4 RF front-ends. We demonstrate the flexibility of WiSCoP by using it to prototype a fully standard compliant IEEE 802.15.4 stack with real-time performance and cross-layer integration. %The demonstration is composed of two showcases which shows WiSCoP flexibility and real-time performance.

\section{Introduction}

Wireless networks have experienced tremendous growth and transformations over the last decade, fueled by the advent of popular devices and services. Future wireless networks will undergo changes in various perspectives and a paradigm shift from the concept of connected people to the concept of connected devices also known as the Internet of Things (IoT). This concept will introduce massive amount of diverse wirelessly connected objects, putting significant pressure on the already limited wireless spectrum. 
More than a decade ago, cross-layer optimization has been introduced as a principle for developing customized protocols that have efficient network resource allocation \cite{shakkottai2003cross}. However, the wireless industry choose to follow a layered design approach, as it allows development of simple, modular and interoperable protocols. This principle causes information hiding between protocol layers, i.e. information about operation at one layer cannot be used by higher or lower layers. Therefore, radio chipset vendors offer limited coordination between physical (PHY), medium access control (MAC), and higher-layer protocols. As a result, PHY and MAC layer innovations are usually prototyped separately without the possibility for joint optimization and real-world evaluation.

\begin{figure}
\centering
\includegraphics[height=2.2in]{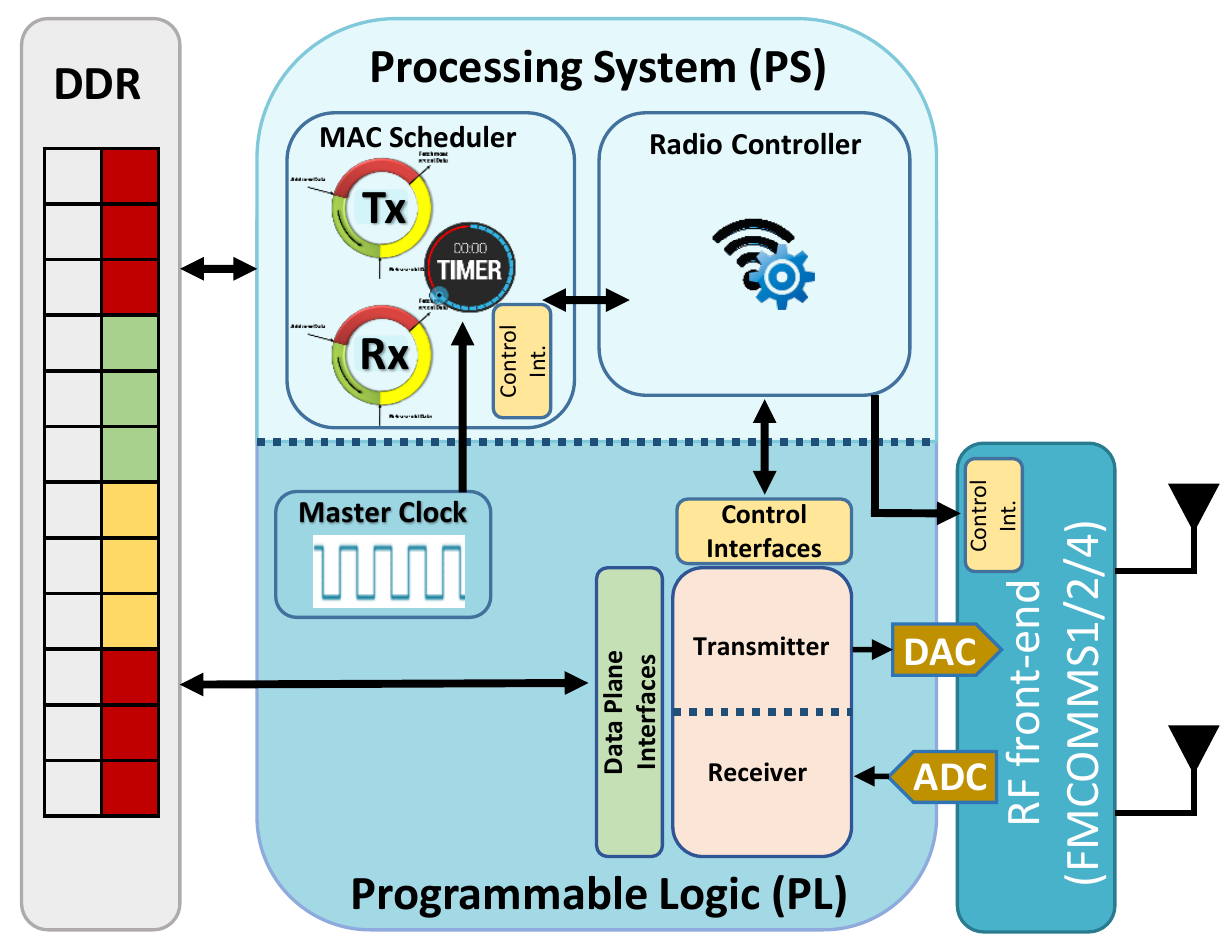}
\caption{WiSCoP architecture showing different modules.}
\label{fig:Fig1}
\end{figure}

Novel PHY algorithms are typically prototyped on SDR platforms. These platforms consist of a RF front-end and a host PC with CPU as main processing unit. However, CPUs cannot guarantee predictable processing latency due to its sequential processing nature. This is the main reason why development of MAC algorithms that control SDR-based PHY layer is extremely difficult. Moreover, such platforms have a large form factor and consume significant power which precludes their distributed deployment. For these reasons, most MAC algorithms are developed and evaluated on embedded platforms. Due to diverse radio chipset capabilities and hardware specifics, evaluating the same MAC layer algorithm on different embedded platforms requires significant development time and effort. Furthermore, benchmarking 
innovative cross-layer PHY/MAC optimized algorithms on diverse platforms is almost impossible.  

To tackle this challenge, we present WiSCoP\footnote{WiSCoP will be available to the research community at https://github.com/TarikKazaz/WiSCoP.}, a novel platform for prototyping cross-layer optimized protocols. Our platform benefits from recent advances in the field of hybrid FPGA technology that tightly couples a FPGA fabric with a hard core CPU on a single die. The very first related work \cite{vermeulen2015demo} introduced a platform for prototyping cross-layer optimized wireless protocols.  However, this platform is closed source, requires expensive LabVIEW license and is based on a host PC and large-scale FPGA which makes it difficult for wide usage and deployment. Compared to existing work \cite{vermeulen2015demo}\cite{bloessl2013gnu} WiSCoP is a flexible, small scale, low power, fully integrated embedded platform with support for real-time experimentation and distributed deployment.

\end{abstract}

%
% NOTE
%
% Do not provide category, terms, keywords for the reviewed submission.
% They will only be added for the camera-ready version. Instructions will
% be provided for the camera ready version.

% \category{H.4}{Information Systems Applications}{Miscellaneous}
%A category including the fourth, optional field follows...
% \category{D.2.8}{Software Engineering}{Metrics}[complexity measures, performance measures]
% \terms{Delphi theory}
% \keywords ACM proceedings, \LaTeX, text tagging}

\section{WiSCoP Architecture}
  \label{sec:intro}

WiSCoP is implemented on top of a Zynq SoC with additional FMCOMMS1/2/4 RF front-ends. The Zynq SoC consists of a dual-core ARM Cortex-A9 processor as the PS unit and an Artix-7 FPGA as the PL unit. The PS and PL are connected through a high-performance (HP) interface and DMA. Apart from the HP interfaces, there are a number of general purpose (GP) interfaces which allow bidirectional register access between the PS and PL. The RF front-ends support frequency range from 60MHz to 6GHz with passband bandwidth of up to 200MHz.

For PHY layer prototyping WiSCoP uses Xilinx HDL or HLS design flows which are available for free through the Xilinx University Program. Prototyping of MAC algorithms can be done in C and compiled for execution on ARM processor. The demo illustrates the real-time performance, IEEE 802.15.4 standard compliance and the flexibility of WiSCoP.
%We illustrate functionalities of WiSCoP by using it to prototype a fully standard compliant IEEE 802.15.4 stack with real-time performance and cross-layer communication. 
The overall hardware design consumes 18\% of HW resources available on Z-7020 Zynq chip. 

The architecture of WiSCoP, shown on Figure \ref{fig:Fig1}, has three modules: \textit{Flexible PHY Layer}, \textit{Medium Access Scheduler} and \textit{Radio Controller}. 

\textbf{Flexible PHY layer.} This module provides a IEEE 802.15.4 2.4 GHz fully standard compliant PHY and CRC engine. It consists of transmit (Tx) and receive (Rx) baseband processing chains. These chains are composed of controllable processing units with exposed read and write registers that support real-time monitoring and control of PHY parameters. Each processing unit is implemented in Verilog and has a standardized AXI interface. The Tx and Rx chains are connected with an ARM processor through a HP interface. This allows high speed data transfer between the PHY and MAC layers. Both, Tx and Rx chains, process samples at 8Msps.  Compared to typical transmit and receive functionalities, our PHY supports additional capabilities:
\begin{itemize}
\item Parametrized control of all signal processing units: Spreading Factor and Pulse Shaping.
\item Customization of PHY layer packet frames including Preamble, SFD and Length.% can be modified to non-standard values.
\item Reporting of PHY parameters to the radio controller.
\end{itemize}

\textbf{Medium access scheduler (MAS).} MAC algorithms usually need to meet rigorous timing constraints. For instance, in IEEE 802.15.4 an acknowledgment packet needs to be sent 192$\mu$s after successful packet reception. Moreover, TDMA-based schemas require precise time-driven packet scheduling. To meet such constraints, we implemented two ring buffers for packet storing and a simple scheduler based on a hardware timer. Ring buffers are directly mapped to shared DDR memory. Data is transferred between shared memory and PHY with support of the DMA controller. This module API provides functions such as: load\_packet, get\_packet, set\_transmission\_time, etc. The hardware timer on the ARM processor ensures real-time execution of these functions. We validated this through a signal analyzer performing power vs. time signal measurements. Results showed that the MAS enables 1$\mu$s packet scheduling accuracy and an average ACK response time of 7$\mu$s. %Compared to off-the-shelf platforms this performance is stable and does not vary with packet size.

\textbf{Radio controller (RC).} The RC is a central control point of WiSCoP. Through the usage of exposed interfaces it configures PHY and RF front-end parameters and coordinates the actions of MAS. It also provides an interface to higher network protocols. Due to the unpredictable and dynamic nature of the wireless medium it is important to trigger radio control actions at precise moments in time or on specific events. This has been solved by adding support for customizable generation of hardware IRQs from the PHY.
%To tackle this challenge, WiSCoP provides customizable generation of hardware IRQs from PHY. %At the moment 
%These IRQs are generated when first sample of Tx waveform is sent to DAC, or when SFD field of packet is received. However, this can be easily changed by marking specific register values.
WiSCoP provides generic interfaces which support execution of custom radio control programs and widely used network stacks such as: ContikiOS or FreeRTOS port of OpenWSN.  

\begin{figure}
\centering
\includegraphics[height=1.925in]{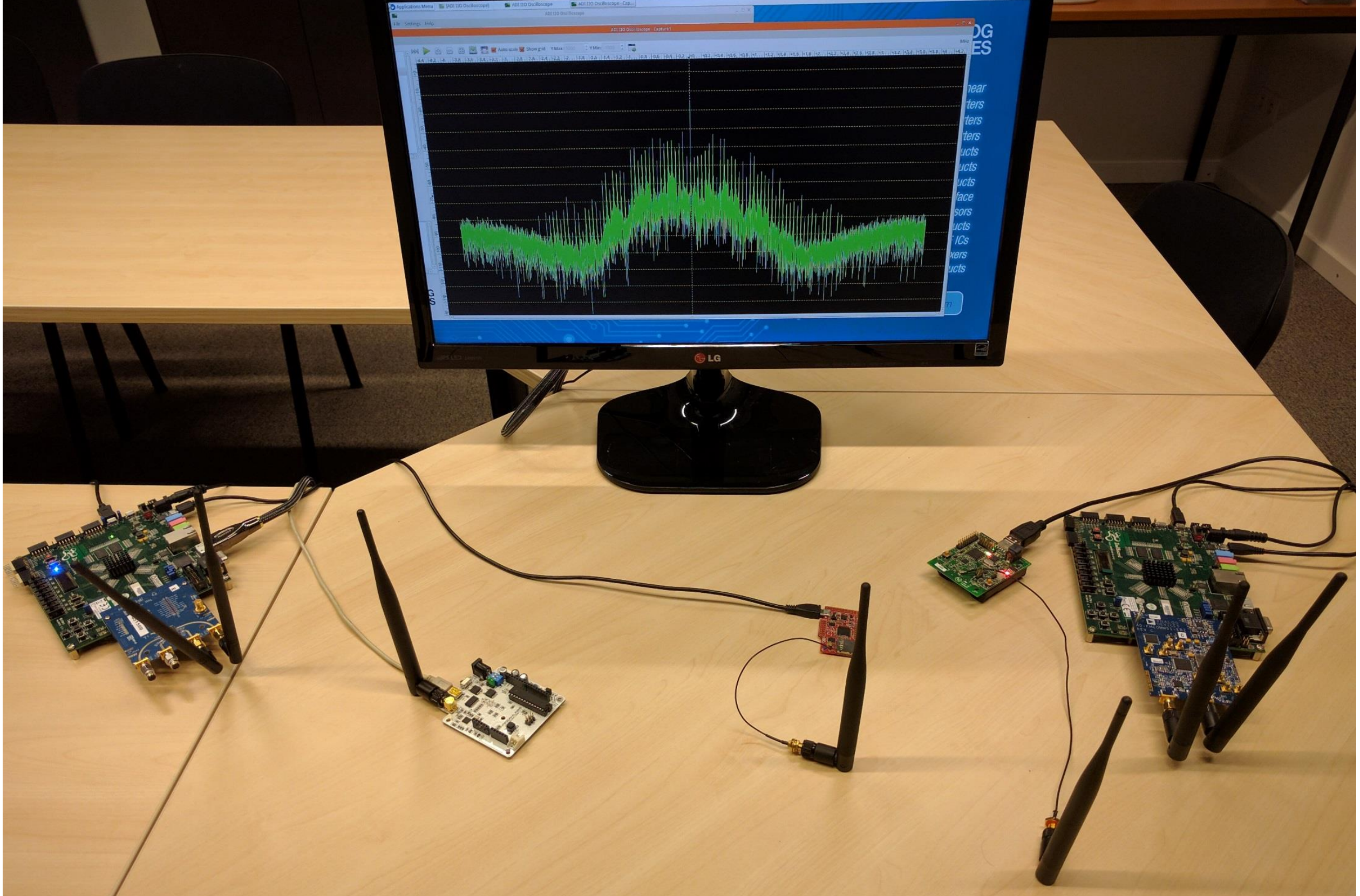}
\label{fig:Fig2}
\caption{Overview of the demo setup.}
\end{figure}

\section{Demonstration}
Figure \ref{fig:Fig2} shows the demo setup for WiSCoP. The setup consist of two 802.15.4 off-the-shelf sensor nodes with different radio chips, two WiSCoP-enabled nodes and one USRP for visualization purposes. The demo has two showcases. The first, demonstrates the compatibility of WiSCoP with two off-the-shelf nodes. During this stage the PHY and MAC parameters that are exposed to the RC will be displayed on the screen. The second, proves the flexibility of our PHY layer by changing PHY parameters during communication between two WiSCoP nodes. During this phase the achieved throughput is visualized. 
%where one transmits a continuous data stream while other one is receiving it. 

%During the demo we will change in real-time spreading factor of PHY layer to increase throughput. The achieved throughput will be displayed on the additional screen. 

%\section{Conclusions}
%This demo presents WiSCoP, an embedded platform for cross-layer prototyping and benchmarking of wireless protocols. Two separate showcases prove its real-time performance, IEEE 802.15.4 standard compliance and flexibility. 

\section{Acknowledgements}
This work has been partially supported by the European Horizon 2020 Programs under grant agreement n$^{\circ}$732174 and n$^{\circ}$645274. %The authors would like to thank Xilinx, Inc. for their donation of Zynq evaluation board.

%
% NOTE
%
%ACKNOWLEDGMENTS: do not provide for submission
%\section{Acknowledgments}
%This section is optional; it is a location for you
%to acknowledge grants, funding, editing assistance and
%what have you.  In the present case, for example, the
%authors would like to thank Gerald Murray of ACM for
%his help in codifying this \textit{Author's Guide}
%and the \textbf{.cls} and \textbf{.tex} files that it describes. 
%Both files are also hacked by John Heidemann and Rasit Eskicioglu.


\balance
\bibliographystyle{unsrt}%{abbrv}
\nocite{*}
\bibliography{sigproc}  % sigproc.bib is the name of the Bibliography in this case
\end{document}